\documentclass[12pt,preprint]{aastex}
\def\kms{{\rm km\,s^{-1}}}

\def\masyr{{\rm mas}\,{\rm yr}^{-1}}

\def\nltt{{\rm NLTT}}
\def\rpm{{\rm RPM}}
\def\lim{{\rm lim}}
\def\det{{\rm det}}
\def\obs{{\rm obs}}
\def\mod{{\rm mod}}
\def\break{{\rm break}}
\def\rpm{{\rm rpm}}
\begin{document}

\title{Stellar Halo Parameters from 4588 Subdwarfs}

\author{Andrew Gould}
\affil{Department of Astronomy, The Ohio State University,
140 W.\ 18th Ave., Columbus, OH 43210}
\authoremail
{gould@astronomy.ohio-state.edu}

\singlespace

\begin{abstract}

Using a reduced proper motion discriminator, I obtain a sample of 
4588 subdwarfs from the Revised NLTT Catalog of Salim \& Gould.
The ample statistics and low contamination permit much
more precise determinations of halo parameters than has previously
been possible.  The stellar halo is not moving with respect to the
Local Standard of Rest (LSR) in either the vertical or radial direction,
up to uncertainties of $2\,\kms$.  This indicates that either the LSR
is on a circular orbit or the Sun happens to lie very close to an
extremum of the LSR's elliptical orbit.  Similarly, tentative detections
of vertical proper motion of Sgr A* relative to the LSR are either
incorrect or they reflect real physical motion of the central black
hole relative to the Galactic potential.  The correlation coefficients
of the halo velocity ellipsoid, which would reflect any possible misalignment
between its principal axes and the cardinal directions of the Galaxy, 
vanish to within 2\%.  The halo subdwarf luminosity function peaks at 
$M_V\sim 10.5$ with a full width half maximum of about 2.5 mag.

\end{abstract}
 
\section{Introduction
\label{sec:intro}}

Samples of nearby halo stars can be analyzed to find the bulk properties
of the population: their velocity, spatial, and metallicity 
distributions, as well as their luminosity function.  The principal
difficulty is obtaining a sample that is large enough to draw statistically
significant conclusions while still not being contaminated with disk and
thick disk stars, which locally outnumber halo stars by a factor $\sim 10^3$.

The most secure method to construct such a sample would be to obtain
parallaxes, proper motions, and radial velocities for a larger, unbiased 
sample of stars and then select halo stars based on their space motions.
The GAIA satellite would be able to do this, but even under the most optimistic
projections, its data will not be available for well over a decade.

In the absence of such ideal datasets, most nearby 
halo samples have been culled
from catalogs of high proper-motion stars (although there are a few
notable exceptions to this rule).  For example, \citet{dahn} obtained
trig parallaxes for about 100 proper-motion selected stars, thereby determining
their transverse velocities, distances, and absolute magnitudes.  By 
rigorous selection on transverse velocity ($v_\perp>260\,\kms$), they obtained
a sample that was virtually free of disk and thick-disk contamination.
The distances and absolute magnitudes then allowed them to measure
the luminosity function (LF).  Of course, to do so they had to correct for
the halo stars that were eliminated from their sample (along with the
unwanted disk stars) by their stringent velocity criterion, and this in turn
required a model of the halo velocity distribution.

The best such model up to that date was constructed by \citet{crb}, who
used maximum likelihood to decompose two proper-motion selected samples 
into disk, thick disk, and halo components making use of both photometric
and proper-motion data.  They thereby identified different populations
within the data, even though individual stars could not generally
be unambiguously associated with a specific population.
In particular, \citet{crb} showed that the likelihood fit was significantly 
improved by allowing for a third ``intermediate'' or thick disk population
rather than just two.  The kinematics of the halo when so fit were more
extreme than in the 2-component fit earlier obtained by \citet{bc}
because thick-disk contamination was drastically reduced.

RR Lyrae stars are halo tracers selected on variability rather than
proper motion.  Estimates of the RR Lyrae absolute magnitude from 
statistical parallax automatically yield the velocity ellipsoid.
While this technique has been applied for almost a century, only in the
last decade or so has it been realized that the RR Lyrae samples 
are actually mixtures of thick-disk and halo stars.  Since statistical
parallax uses both radial velocities (whose spectra also yield metallicity
information) and proper motions, and since it derives distances for all stars,
full kinematic as well as metallicity information is generally available.
In a series of papers, \citet{layden94,layden95,layden97}
both systematized pre-existing data and obtained
substantial new data, thereby laying the basis for a new
statistical parallax solution that clearly separated the thick-disk and
halo populations using a combination of kinematic and metallicity criteria
\citep{layden}.  \citet{pg1,pg2} and \citet{gp} (collectively PG$^3$)
introduced new mathematical
methods and on this basis conducted a thorough overhaul of the 
\citet{layden} sample, recalibrating much of the old photographic photometry,
incorporating more modern extinctions, identifying suspicious astrometry,
and developing a new method to incorporate non-RR-Lyrae radial velocities
into the analysis.  

Of particular note in the present context, PG$^3$
were the first to measure five of the nine components of the halo
velocity ellipsoid: all previous analyses had measured the three
diagonal components of the velocity dispersion tensor and the component
of bulk motion in the tangential direction (the asymmetric drift),
but had assumed that the off-diagonal components as well as the bulk motion
in the radial and vertical directions were zero.  While the PG$^3$ measurements
all turned out to be consistent with zero (in the frame of the Local
Standard of Rest -- LSR), the error bars were tantalizingly close to being
able to probe some interesting scientific questions.

The bulk motion of the halo in the radial and vertical directions is more 
likely to coincide with the rest frame of the Galaxy than is the motion of
the LSR.  The LSR could well be on an elliptical orbit, in which case
it would be moving towards or away from the Galacitic center unless
the Sun happened to lie at an extremum of this orbit.  Indeed, \citet{bs}
claimed that the LSR is moving outward at $14\,\kms$ based on radial-velocity
measurements of gas in the outer galaxy (assumed to be on circular orbits).  
On the other hand, \cite{ms} concluded that the LSR was moving inward at
$6.6\pm 1.7\,\kms$ based on radial velocities of carbon stars in the outer 
Galaxy.

Similarly, if the Milky Way disk is warped, then one would expect the LSR to
be moving either up or down relative to the Galactic rest frame, unless the 
Sun happened to be at an extremum of the warp.  \citet{backer} found that
Sgr A* is moving down at $17\pm 6\,\kms$ relative to the LSR.  If 
the supermassive
black hole associated with Sgr A* is assumed to be at rest with respect to
the Galaxy, then this apparent motion would actually be a reflex of the
warped motion of the LSR.  On the other hand, \citet{reid} 
find that Sgr A* is moving in the opposite direction (although with much 
larger errors) at $15\pm 11\,\kms$.  New more
precisements measurements are expected soon (M.\ Reid, private communication
2001).

For a roughly isotropic ensemble of $N_s$ stars, the bulk motion $U_i$ can be 
measured with a precision,
\begin{equation}
\sigma(U_i) \sim \sqrt{3\,c_{ii}\over n_d N_s}
\label{eqn:sigmaui}
\end{equation}
where $c_{ii}$ is the dispersion in the $i$th directin and $n_d$ is
the number of components of the velocity measured for each star.
For the PG$^3$ sample, $N_s\sim 170$ and $n_d=3$, while
$c_{11}\sim (160\,\kms)^2$ and $c_{33}\sim (90\,\kms)^2$.  Hence
$\sigma(U_1) \sim 13\,\kms$ and $\sigma(U_3)\sim 8\,\kms$.  The PG$^3$ 
measurement errors were therefore not quite small enough to probe these 
interesting questions.

Of course, measurement of a difference between the bulk-halo and LSR 
velocities would not be unambiguous evidence of LSR motion \citep{oresme}.
For example, the angular momentum vector of material infalling onto the
Milky Way could have radically changed between the time of the formation
of the halo and disk.  The former therefore could in principle be
rotating in a basically polar orbit (albeit with low Mach number) 
relative to the latter.  Hence any sort of relative motion would be
intriguing evidence of a non-simple Galaxy whose exact origins would
have to be sorted out making use of other data and arguments.

Similarly, the off-diagonal elements of the velocity dispersion tensor
(normalized to the diagonal elements) $\tilde c_{ij}$ could potentially
provide evidence of asymmetries of the Galaxy that would reflect on its
origins.  The errors in these quantities are $\sim (n_d N_s/3)^{-1/2}$,
or about 8\% for the RR Lyrae sample.  To the best of my knowledge,
no one has investigated what might cause these quantities to differ
from zero, so I do not know whether their consistency with zero at the
8\% level challenges or confirms any theory.  Nevertheless, it seems
interesting to try to probe the off-diagonal elements at higher precision.

The status of the LF and local density of the stellar
halo are also somewhat controversial.  \citet{dahn} find that the LF
peaks at around $M_V\sim 12$, in qualitative agreement with the shape of
the LF seen in undisturbed globular clusters \citep{piotto}.  However,
\citet{bc} and \citet{gfb} find a roughly flat LF over the interval
$9\la M_V \la 13$.  While \citet{dahn} and \citet{bc} both studied
local stars drawn from proper-motion catalogs, \citet{gfb} adopted a
radically different approach: they located stars in {\it Hubble Space
Telescope} images that were too faint at their observed color to be in
the disk and so were assigned absolute magnitudes and distances based on
a halo color-magnitude relation.  These stars were generally quite
distant ($\ga 3\,$kpc) and therefore perhaps not directly comparable
to the local samples.  \citet{sz} had earlier suggested that the stellar
halo actually has two components, one roughly spheroidal and one highly
flattened.  (The highly flattened component is not to be confused
with the thick disk: it is not rotating significantly.)\ \ In their model, the 
two components have roughly equal densities at the solar circle.  Such a model
predicts that the halo density should be roughly twice as great in the
solar neighborhood as it is at a similar Galactocentric radius but a few
kpc above the Galactic plane.  Indeed, \citet{gfb} found a halo density
that was lower than the \citet{dahn} measurements by just this fraction.  
Recently \citet{siegel} have argued on the basis of a sample of 70,000
stars along multiple pencil beams that even a 2-component halo model
is inadequate to explain their star counts.

The newly released
Revised NLTT Catalog \citep{bright,faint} allows one to obtain a very large
and very clean sample of halo stars.  The reduced proper-motion (RPM)
diagram using the newly obtained $V-J$ colors clearly separates main-sequence
stars, subdwarfs, and white dwarfs into different tracks \citep{rpm}
in sharp contrast to RPM diagram 
constructed from the original NLTT \citep{luy}.
Although the first release of this catalog covers only 44\% of the sky,
it contains more than 5000 local
halo stars, well over an order of magnitude more 
than have ever been cleanly distinguished from disk stars on a star-by-star
basis.  This sample therefore opens the way to a much more detailed
study of the local halo population than has previously been possible.
In its present form, the sample does have some limitations.  
Since most of its stars lack radial velocities
and parallaxes, it is not possible to establish the absolute distances
or the amplitude of the velocity ellipsoid based on the Revised NLTT 
Catalog alone.
Nevertheless, the amplitude of the velocity ellipsoid is already 
known with a precision of about 10\% from previous studies, 
and by incorporating this
external information one can obtain much more precise measurements of
the five components of the ellipsoid that are currently poorly measured:
$U_1$, $U_3$ and $\tilde c_{ij}$.  Once the velocity scale is set, the
mean distances to the stars are also determined, which permits one to
measure the LF.  The large number of stars in the sample therefore offers
the hope of probing the bottom of the subdwarf sequence which, because
of its dimness, is poorly represented in magnitude limited samples.
Finally, the catalog contains a large number of stars from the Galactic
plane to about $z\sim \bar V_\perp/\mu_{\rm lim}\sim 350\,$pc
above the plane, where $\bar V_\perp\sim 300\,\kms$ is the typical
transverse speed seen toward the Galactic poles and 
$\mu_{\rm lim}= 180\,\masyr$ is the proper-motion limit of the catalog.
While this distance is short compared to the several kpc's hypothesized
as the height of the flattened halo component, the large number of
stars in the sample may yield a statistically significant statement
about the presence of a density gradient on these larger scales.

Maximum-likelihood (ML) analysis
is absolutely critical for extracting halo parameters
from this catalog.  For example, since the mean tangential velocity of stars
seen toward the Galactic poles is $\sim 200\,\kms$, one might naively
expect that the stars selected toward the poles would have, on average, this
velocity.  However, given the fact that the sample is proper-motion limited,
for all but the dimmest absolute magnitudes the number of stars seen
with velocities that are $1\,\sigma$ higher than average 
($300\,\kms$) is $(300/100)^3=27$
times higher than the number with velocities that are $1\,\sigma$ lower
($100\,\kms$).  This severe selection bias does not {\it directly} affect any
other parameters.  However, it couples through the highly uneven (but
perfectly known) sky coverage from 2MASS, 
to {\it indirectly} affect essentially
all other parameters.  These effects can only be removed by comparing
the predictions of models with the observations, as ML does automatically.

Hence, I begin in \S~\ref{sec:maxlike} by giving a careful summary of
the ML modeling procedure.  In \S~\ref{sec:results},
I present my results and comment on various aspects of these whose
interpretation requires caution.  Finally, in \S~\ref{sec:discuss}, I
compare my results to previous work and briefly discuss the implications
of this comparison.  I reserve to the Appendix a somewhat technical 
discussion of the problems in determining the completeness of the Revised
NLTT Catalog and the impact of this completeness on parameter estimation.

\section{Maximum Likelihood Formulation
\label{sec:maxlike}}
\subsection{General Equation
\label{sec:genmaxlike}}
I use ML to estimate the parameters of the stellar halo.  
In general,
a given dataset is described by $m$ observables $z_{\obs,i}$, which I
collectively denote $z^m_\obs$.  If this $m$-dimensional space of observables
is divided into bins of volume $\prod_{i=1}^m\Delta z_{\obs,i}$, then the
likelihood  of detecting $n_k$ objects in the $k$th bin is
${\cal L}_k = \tau_k^{n_k}\exp(-\tau_k)/n_k!$, where 
$\tau_k = P_k(z^m_\obs)\prod_{i=1}^m
\Delta z_{\obs,i}$, and $P_k$ is the probability
density predicted by a given model.  If the bins are now made sufficiently
small that $\tau_k\ll 1$, then $n_k\le 1$ and hence $n_k!=1$.  The logarithm
of the product of the likelihoods from all the bins is therefore,
\begin{equation}
  \ln {\cal L} = \sum_k \ln {\cal L}_k = \sum_k n_k \ln \tau_k -\sum_k \tau_k.
\label{eqn:lnlike1}
\end{equation}
The last term is just the total number of detections expected in the model,
$N_{\rm exp}$.  Since the $n_k$ in the first term are either 0 or 1, 
equation (\ref{eqn:lnlike1}) can be rewritten,
\begin{equation}
  \ln {\cal L} = \sum_{k=1}^{N_\det} \ln [P_k(z^m_\obs)\prod_{i=1}^m
\Delta z_{\obs,i}]  - N_{\rm exp}.
\label{eqn:lnlike2}
\end{equation}
where $N_\det$ is the total number of detections.
In general, however, the probability density is not most naturally written
directly as a function of the observables $z_{\obs,i}$, but rather of
the model coordinates, $z_{\mod,i}$, which are evaluated at the observables.
Equation (\ref{eqn:lnlike2}) can be rewritten in terms of these,
\begin{equation}
  \ln {\cal L} = \sum_{k=1}^{N_\det} \ln \{P_k[z^m_\mod(z^m_\obs)]{\cal J}\}
 -N_{\rm exp} + N_\det\sum_{i=1}^m\ln \Delta z_{\obs,i}.
\label{eqn:lnlike3}
\end{equation}
where ${\cal J}$ is the Jacobian of the transformation from the observables
to the model coordinates.
For a given set of observations, the last term is the same for all models 
and so can be dropped.

\subsection{Jacobian
\label{sec:jacobian}}

In the present case, for each star
there are six observables: the angular position on
the sky $(l,b)$, the proper motion $(\mu_l,\mu_b)$, and the two photometric
magnitudes $(V,J)$.  There are also six model coordinates: the three spatial
coordinates $\bf r$, the two components of transverse velocity $\bf v_\perp$, 
and the absolute magnitude, $M_V$.  Hence,
\begin{equation}
{\cal J} = \bigg|{\partial ({\bf r},{\bf v_\perp},M_V)\over
\partial(l,b,\mu_l,\mu_b,V,J)}\bigg| 
= r^4\cos b\bigg|{\partial(r,M_V)\over\partial(V,J)}\bigg|.
\label{eqn:jac1}
\end{equation}
To evaluate ${\cal J}$, I write $r$ and $M_V$ as implicit functions
of $V$ and $J$, making use of the
color-magnitude relation, $M_V = F[(V-J)_0]$.  
\begin{equation}
r = 10^{0.2[V - A_V(r) - M_V] + 1},\quad
M_V =  F\biggl[V-J-{A_V(r)\over R_{VJ}}\biggr],
\label{eqn:disrel}
\end{equation}
Here, $A_V(r)$ is the extinction along the line of sight and 
$R_{VJ}=1.38$ is the ratio of total to selective extinction.
Partial differentiation of equation (\ref{eqn:disrel}) yields the
matrix equation,
\begin{equation}
\biggl(\matrix{ r & {5\over \ln 10}+ A' r\cr
1 & {F'A'\over R_{VJ}}\cr} \biggr)
\biggl(\matrix{{\partial r\over \partial V} &{\partial r\over \partial J} \cr
{\partial M_V\over \partial V} & {\partial M_V\over \partial J}\cr}\biggr)
=\biggl(\matrix{ r & 0\cr F' & -F'\cr}\biggr)
\label{eqn:mateq}
\end{equation}
whose determinant gives,
\begin{equation}
{\cal J} =  {\ln 10\over 5}\cos b\, r^5 F'Q,\quad
Q\equiv \bigg| 1 - {\ln 10\over 5}{d A_V\over d\ln r}
\biggl({F'\over R_{VJ}}-1\biggr)\bigg|^{-1}.
\label{eqn:jacfull}
\end{equation}
If the reddening vector were parallel to the subdwarf sequence 
$(F'/R_{VJ}\simeq 1)$, then the additional term $Q$
in equation (\ref{eqn:jacfull})
would be negligible.  In fact, however, this ratio is roughly
$F'/R_{VJ}\sim 2.6$, which means that at typical distances $r\sim 300\,$pc
and low Galactic latitudes, $\ln Q\sim 0.12$, and so it cannot be ignored.
Note that the prefactor $(0.2\ln 10\cos b)$ is an irrelevant constant
and can be dropped in practical calculations, so that 
${\cal J}\rightarrow r^5 F'Q$.

\subsection{Model Parameters
\label{sec:modparms}}

I model the stellar halo distribution as the product of a luminosity 
function (LF), a velocity distribution, and a density profile.  
In addition, halo stars are assumed to obey a linear color-magnitude
relation 
\begin{equation}
M_V = F[(V-J)_0] = a(V-J)_0 + b.  
\label{eqn:cmr}
\end{equation}
In principle, one might
also assume that this relation has some intrinsic dispersion.  However,
for reasons that I discuss below, I do not include such a parameter
in the model.  Finally, not all halo stars satisfying the selection
criteria will be detected.  I therefore include in the model two parameters
$(V_\break,f_\break)$ that describe the completeness as a function of 
apparent $V$ magnitude.
$$C(V) = 1\,\quad (V<12),\qquad$$
$$C(V) = {(V_\break - V) + f_\break(V-12)\over (V_\break - 12)} 
\,\quad (12<V<V_\break),\qquad$$
\begin{equation}
C(V) = {f_\break(20 - V_\break) + (V-V_\break)\over (20 - V_\break)} 
\,\quad (V_\break<V<20)
\label{eqn:complete}
\end{equation}
This form is motivated by the fact that NLTT is known to be complete to at 
least $V=11$ away from the plane, and the Revised Catalog has captured
essentially all of these stars \citep{bright}.  The completeness falls
precipitously at faint magnitudes, $V\ga 18$.  The simplest hypothesis
is that it is linear over the intervening magnitudes.

I model the LF with 13 free parameters, one for each magnitude bin
centered at $M_V=3$ to $M_V=15$.  I model the velocity distribution
as a Gaussian  ellipsoid with 9 parameters: three for the bulk motion,
$U_i$, three for the diagonal components of the dispersion tensor,
$c_{ii}$, and three for the off-diagonal components $c_{ij}$ $(i<j)$.
In practice, I use the normalized components of the latter (the
correlation coefficients) $\tilde c_{ij} = c_{ij}/(c_{ii} c_{jj})^{1/2}$.
It is known that the halo velocity distribution is highly non-Gaussian,
with a kurtosis that is higher than Gaussian in the vertical direction,
lower than Gaussian in the radial direction, and roughly Gaussian in
the direction of rotation (PG$^3$).  However, modeling the
non-Gaussian character of the distribution would be quite complicated, and
it is straight forward to show that a ML fit of a non-Gaussian distribution
to a Gaussian model returns unbiased estimates of the first two moments.
Hence there is no benefit to modelling the non-Gaussian form of the 
distribution unless one wants to investigate the higher moments of
the distribution.  Since these are not a focus of interest in the current
paper, I opt for the simpler Gaussian model.  

I model the halo distribution as falling as a power law
with Galactocentric distance $R$ and exponentially with distance
from the Galactic plane $z$, i.e. $\rho = \rho_0(R/R_0)^{-\nu}
\exp(-\kappa |z|)$.  The spatial distribution is therefore described by two 
parameters,  $\nu$ and $\kappa$, where $\kappa$ may be regarded as the
inverse scale height.  I adopt $R_0=8\,$kpc.  

Hence, I begin with 28 free parameters, 13 for the LF, 9 for the
velocity ellipsoid, 2 for the color-magnitude relation, 2 for the
completeness fucntion, and 2 for the density profile.  However, as I
now explain, there is one almost perfect degeneracy among these parameters,
and therefore one of them must be fixed.  If the zero-point $b$ of the 
color-magnitude relation (eq.\ [\ref{eqn:cmr}]) is increased by $\Delta b$,
and all the bulk velocities $U_i$ and dispersions $\sqrt{c_{ii}}$ are
reduced by $10^{-0.2\Delta b}$, then all of the model's predicted proper 
motions will remain unchanged.  The only difference will be that a star's
absolute magnitude (inferred from its color) will be increased by $\Delta M_V =
\Delta b$, and so the inferred density of stars of each resulting magnitude
bin will be increased by $10^{0.6\Delta b}$.  Actually, this scaling 
remains perfect only in the limit $A_V\rightarrow 0$, but since the extinction
is quite small, the degeneracy is almost perfect.

I therefore fix 
\begin{equation}
U_2 = -216.6\,\kms,
\label{eqn:gpu2}
\end{equation}
the value measured by \citet{gp} for their ``kinematically selected'' sample
of halo RR Lyrae stars. \citet{gp} also evaluated the velocity ellipsoid
for a ``non-kinematically selected'' sample of halo stars, the magnitude
of whose $U_2$ component is somewhat smaller than given by equation
(\ref{eqn:gpu2}).  However, as I describe in \S~\ref{sec:selcrit} below,
the present sample is effectively selected using a combination of
kinematic and metallicty criteria, just as was true for the \citet{gp}
``kinematically selected'' sample.  Therefore, the scale of the
velocity ellipsoid should be fixed by the ``kinmatically selected'' 
rather than the 
\cite{gp} ``non-kinematically selected'' sample.

Finally, as noted above, I do not include a parameter for the dispersion
in the color-magnitude relation despite the fact that the halo is known
to contain a range of metallicities, and therefore a range of absolute
magnitudes at fixed color, and hence some dispersion $\sigma(M_V)$.  Since
I do not include this term, the velocity dispersions found by the 
ML fit will be larger than the true dispersions by 
\begin{equation}
\Delta c_{ii} = (U_i^2 + c_{ii})[0.2\ln 10\sigma(M_V)]^2.
\label{eqn:ciichange}
\end{equation}
Hence, in principle, if one knew the $c_{ii}$ sufficiently well, one
could fix them (or one of them or their sum), and fit for $\sigma(M_V)$.
However, as I will show in \S~\ref{sec:results}, 
the differences $\Delta c_{ii}$ are
smaller than the present uncertainties in the $c_{ii}$, and therefore
this is not a practical possibility.  Thus, the fit parameters that I am
calling ``$c_{ii}$'' are actually shorthand labels for
$c_{ii} + \Delta c_{ii}$.  This means that the derived parameters will not 
yield new determinations for the $c_{ii}$.  The best one can do is use these
measurements to place rough upper limits on $\sigma(M_V)$.
That is, among the nine velocity-ellipsoid parameters $U_i$, $c_{ii}$,
$\tilde c_{ij}$, new values will be obtained for only five: $U_1$, $U_3$,
and $\tilde c_{ij}$.

\subsection{Data Characteristics
\label{datachar}}

Among the six observables $(l,b,\mu_l,\mu_b,V,J)$ only the $V$ magnitude
has significant errors.  The angular coordinates $(l,b)$ are known to
$<100\,$mas, about 6 orders of magnitude smaller than the scale on which
there are significant density gradients.  The proper-motion errors are
$5.5\,\masyr$ or 3\% of the NLTT proper-motion threshold.  Since the
intrinsic dispersion in proper motions is of order unity, and since
the measurement errors add in quadrature to these, these errors 
are utterly negligible.
The $V$ errors are about 0.25 mag \citep{nearbylens}.  As discussed
by \citet{faint}, these are multiplied by 2.1 when entering the RPM and
therefore cannot be ignored.  Finally, the $J$ errors are typically
0.03 mag.  Even though they enter the RPM with somewhat higher weight
(3.1), they are then added in quadrature to the $V$ errors, and so are
also negligible.

The errors in the original NLTT proper motions were $20\,\masyr$
\citep{faint}.  These proper motions are not used directly in the
evaluation of the likelihood, but they do have an indirect effect because
they influenced Luyten's determination of which stars met his proper-motion
threshold of $180\,\masyr$ and so which ultimately entered the Revised
NLTT Catalog \citep{bright,faint}.

\subsection{Selection Criteria
\label{sec:selcrit}}

One wishes to select as large a sample of halo stars as possible, while
effectively excluding stars from other populations.
Moreover, one would like to restrict selection to those areas of the sky with
homogeneous completeness characteristics.  To achieve the first goal, 
I make use of the discriminator $\eta$ introduced by \citet{faint},
\begin{equation}
\eta(V_\rpm,V-J,\sin b) =  V_\rpm - 3.1(V-J) - 1.47 |\sin b| - 2.73.
\label{eqn:eta}
\end{equation}
where $V_\rpm = V + 5\log(\mu)$ is the RPM.
Using the $(V,V-J)$ RPM diagram, \citet{faint} showed that stars in the range
$0<\eta<5.15$ are mostly halo stars.  To be conservative, I restrict the
selection by an additional magnitude on each end and require $1<\eta<4.15$
(see Fig.\ 3 from \citealt{faint}).  Completeness of the original NLTT
catalog deteriorates significantly in areas south of POSS I, and \citet{faint} 
did not even attempt to recover faint NLTT stars in this region because their
method cannot be applied there.  
I therefore require $\delta > -32.\hskip -2pt ^\circ 4$.
\citet{faint} showed that in the Galactic latitude interval $-0.2<\sin b<0.3$,
NLTT completeness of main-sequence stars is severely affected, dropping from
neighboring zones at higher latitude
by a factor $\sim 10.$  While the effect is much smaller
for subdwarfs (and perhaps negligible for white dwarfs), to be conservative,
I restrict selection to stars outside this range.
Finally, of course, the identifications by \citet{faint} rely
critically on 2MASS, 
and so have only been carried out for the 47\% of the sky (57\% of the region 
$\delta > -32.\hskip -2pt ^\circ 4$) that is covered by the second incremental
2MASS release.  Hence, the spatial selection function alone is quite
complex.  This fact, together with the large number of observables,
implies that great care is required to evaluate the likelihood function.

Note that the discriminator $\eta$ (eq.\ [\ref{eqn:eta}]) is effectively
a function of both kinematics and metallicity.  That is, $\eta$ increases
both with higher transverse velocity and with lower metallicity (and so lower
luminosity at fixed color).  It is for this reason that I said 
in \S~\ref{sec:modparms} that the sample is selected using a combination of
kinematic and metallicity criteria.

\subsection{Likelihood Evaluation
\label{sec:evaluation}}

In order to find the model parameters that maximize the likelihood, one must 
compare in a consistent way the likelihood of observing the data given 
different sets of model parameters.  This statement is so
obvious that it would appear not worth mentioning.  However, achieving
such consistency is by no means trivial.

The first term in equation (\ref{eqn:lnlike3}) is relatively straight
forward to calculate because it depends only on differential probability
functions that are multiplied together. However, the second term, the total 
number of stars expected to enter the sample for a given model is quite
complicated in several respects.  First, as discussed in \S~\ref{sec:selcrit}
the selection criteria themselves are complex.  Second, to estimate 
$N_{\rm exp}$ requires an integral over nine dimensions: six for the model
coordinates $({\bf r},{\bf v_\perp},M_V)$, plus three for the measurement 
errors $(V,\mu_{l,\nltt},\mu_{b,\nltt})$.  Recall that even though I am
not making use of the NLTT proper-motion measurements, they still enter
the likelihood function because they affect the sample selection.

Integration over more than four dimensions is in general more efficiently 
carried out by Monte Carlo than directly.  In the present 
case, the complexity of
the 2MASS coverage further reinforces the advantages of Monte Carlo 
integration.  However, such an approach poses significant difficulties
when comparing likelihood estimates at different locations in parameter 
space:  Monte Carlo integration introduces Poisson fluctuations into
the evaluation of $N_{\rm exp}$ which are of order the square root of the
size of the random sample.  While these fluctuations can to some extent be
suppressed by insisting that all realizations have the same sample size,
the induced fluctuations remain of the same order.  To be certain that
these do not induce roughness in the ML surface of
order unity would require Monte Carlo samples of ${\cal O}(10^7)$, which
would be computationally prohibiive.

To counter this problem, rather than directly assembling a separate catalog of
fake stars for each model, I assemble a single catalog of fake stars and
assign the stars different weights according to the model.  That is, I first
choose a baseline model that is reasonably close to the final model.
For each magnitude bin, I assign an absolute magnitude drawn uniformly over
this bin, a physical location 
drawn uniformly from the volume within 1 kpc of the Sun,
and a transverse velocity drawn randomly from the 2-dimensional Gaussian
projected-velocity distribution expected in the baseline model.  I note
the Gaussian probability of each such fake star, but do not
at this point make use of it.

Next I determine the observational characteristics of the star.  For
example, I use the color-magnitude relation of the baseline model and the 
star's distance to obtain the true $(V-J)_0$ color and $V_0$ magnitude.  I 
draw the error in the observed $V$ magnitude from a Gaussian distribution, and 
redden both the color and magnitude according to
a simplified extinction law,
\begin{equation}
A_V(r,b) = 0.075\,|\csc b| [1 - \exp(-r\sin |b|/h_d)]
\label{eqn:av}
\end{equation}
where $h_d=130\,$pc is the dust scale height.  Similarly, I obtain the true
proper motion from the distance and transverse velocity and draw the two 
NLTT proper-motion errors from a Gaussian distribution.  
If the NLTT proper motion exceeds $180\,\masyr$, and the RPM discriminator 
$\eta$ lies within somewhat expanded bounds $(-1<\eta<6.15)$ as
calculated within the baseline model, I accept the star into a master list
of fake stars.  For each model I examine each star on this master list and
recalculate $(V-J)$ using the model's color-magnitude relation.  I accept only
stars satisfying the selection criterion $(1<\eta< 4.15)$ as calculated within
the model.  Next, I determine
$N_{\rm exp}$ by counting all of the fake stars thus accepted and 
assigning each one a
weight that depends on the model.
The weight is a product of factors: $(R/R_0)^{-\nu}$ for proximity
to the Galactic center, $\exp(-\kappa |z|)$ 
for distance from the Galactic plane,
$C(V)$ to take account of completeness, as well as a factor for the
luminosity function of the star's $M_V$ bin.  In particular, I evaluate
the probability of the transverse velocity given the model and divide this
by the tabulated probability of the baeline model.  In this way, 
$N_{\rm exp}$ is evaluated stochastically for an ensemble of models, 
without introducing random noise into the {\it relative} values obtained
for different models.

For purposes of finding the best-fit model, I use a catalog of fake
stars drawn from a model that is 100 times denser than the actual stellar 
halo, so that
the stochastic character of the fake catalog introduces errors in the 
parameter estimates that are 10 times smaller than the Poisson errors.
I use bootstrap to calculate the errors in the parameters, i.e., I
evaluate the scatter in model fits to 25 realizations of the data
formed by drawing randomly from the actual data with replacement.
Each realization is tested against the {\it same} catalog of fake
data.  For this purpose, I use a fake catalog drawn from a model
that is 10 times denser than the actual stellar halo.

\section{Results
\label{sec:results}}

	The best-fit model to the 4588 subdwarfs selected according to
the criteria described in \S~\ref{sec:selcrit} has the following 
characteristics.

\subsection{Velocity Ellipsoid Parameters
\label{sec:velparms}}

	For the bulk halo motion relative to the Sun, I find
\begin{equation}
U_1 = 11.4\pm 2.2\,\kms,\qquad U_3=-5.4\pm 2.4\,\kms,
\label{eqn:u1u3}
\end{equation}
in the (outward) radial and (upward) vertical directions.  Since
the Sun moves relative to the LSR at $-10.0\pm 0.4\,\kms$ and 
$7.2\pm 0.4\,\kms$ in these directions \citep{db},  
this implies that the LSR is moving
relative to the halo at $-1.4\pm 2.2\,\kms$ radially and $-1.8\pm 2.4\,\kms$ 
vertically.  That is, both components are consistent with zero.  If the
halo is assumed to be stationary in both directions relative to the Galactic
potential, then either the deviations of the LSR from a circular orbit must 
be very small, or the Sun must lie close to the extrema of these deviations.
On the other hand, if the halo is not stationary, then it just happens to
have almost exactly the same motion as the LSR, which would be a most 
surprising coincidence.

	The errors in equation (\ref{eqn:u1u3}) include only the statistical
errors within the fit and not the systematic errors induced by fixing
the amplitude of the velocity ellipsoid using the \citet{gp} value for
$U_2=-216.6\,\kms$.  However, as I now show, this systematic error is 
relatively small.  First, the statistical error of $U_2$ is $12.5\,\kms$
or 6\%.  This induces a systematic error in $U_1$ also of 6\%, that is,
$0.7\,\kms$ which is small compared to the statistical error.  There is
a second source of error because, while both the present sample and the
\citet{gp} sample were chosen based on a combination of kinematic and
metallicity criteria, those criteria are not identical, nor even easily
comparable.  Hence, the values of $U_2$ for the two samples need not
be identical.  It is difficult to judge the size of this systematic error,
but it is probably also of order $10\,\kms$, i.e., about 5\%, and therefore
again much smaller than the statistical error.  The systematic errors
scale with $U_i$ and therefore are about half as big for the vertical
motion as the radial motion.

	The three off-diagonal components to the velocity-dispersion 
tensor are
\begin{equation}
\tilde c_{12}= 0.024\pm 0.014,\qquad
\tilde c_{13}= 0.005\pm 0.023,\qquad
\tilde c_{23}= -0.004\pm 0.026.
\label{eqn:tildec}
\end{equation}
That is, all three are consistent with zero at about the 2\% level.
(Because there are five velocity-ellipsoid parameters being fit, the
$1.7\,\sigma$ ``detection'' of $\tilde c_{12}$ cannot be regarded
as even marginally significant.)

	Finally, the three diagonal components are
\begin{equation}
\sqrt{c_{ii} + \Delta c_{ii}} = (162.4\pm 1.4,\ 105.8\pm 1.7,\ 89.4\pm 1.9)
\,\kms,
\label{eqn:cii}
\end{equation}
where $\Delta c_{ii}$ is defined by equation (\ref{eqn:ciichange}).
This can be compared with values of $c_{ii}$ found by \citet{gp}
for kinematically selected RR Lyrae stars.
\begin{equation}
\sqrt{c_{ii}} = (171\pm 10,\ 99\pm  8,\  
90\pm  7)\,\kms,\qquad \rm (RR\ Lyraes),
\label{eqn:rrlyraes}
\end{equation}
If the errors in equation (\ref{eqn:rrlyraes}) were sufficient small,
it would be possible to determine the $\Delta c_{ii}$ in equation 
(\ref{eqn:cii}) and so characterize $\sigma(M_V)$
(the scatter in $M_V$ at fixed $V-J$ color).  See equation 
(\ref{eqn:ciichange}).
However, given the errors, it is immediately clear that $\sigma(M_V)$
is consistent with zero.  To find out what upper bound can be put on
$\sigma(M_V)$, I first note that because $c_{22}/U_2^2\sim 5$, most of 
the potential information comes from the tangential component.  At
the $1\,\sigma$ level, $(\Delta c_{22})^{1/2} < 55\,\kms$.  Hence,
from equation (\ref{eqn:ciichange}), $\sigma(M_V)<0.5$.  This is not
a very interesting $1\,\sigma$ limit.  
Moreover, I have not yet incorporated 
the statistical or systematic uncertainties in the amplitude of the 
velocity ellipsoid as discussed following equation (\ref{eqn:u1u3}).
Hence, equations (\ref{eqn:cii}) and (\ref{eqn:rrlyraes}) present
a reasonably consistent picture, but do not significantly constrain
$\sigma(M_V)$.  I do note, however, that
the comparison of these two equations shows that the choice for
normalizing the velocity ellipsoid, $U_2=-216.6\,\kms$, cannot be off
by more than about 15\%.  If it were, then the fit values for
$(c_{ii} + \Delta c_{ii})^{1/2}$ would also change by 15\%, and these
would then be inconsistent at high significance with the $c_{11}$ and 
$c_{33}$ as measured for RR Lyrae stars.

\subsection{Halo Profile Parameters
\label{sec:profile}}

	The halo density is not expected to vary much over the small
($\sim 300\,$pc) volume that is being probed.  As discussed in 
\S~\ref{sec:modparms}, I therefore model the density profile simply as,
$\rho = \rho_0(R/R_0)^{-\nu} \exp(-\kappa|z|)$.  I find,
\begin{equation}
\nu = 3.1\pm 1.0,\qquad \kappa = 0.022\pm 0.057\,{\rm kpc}^{-1}.
\label{eqn:nukappa}
\end{equation}
The estimate of $\nu$ is consistent with many previous determinations
which, because they are measured over longer baselines, have much small
errors.  For example, \citet{gfb} find $\nu=2.96\pm 0.27$.  The $\kappa$
measurement is quite interesting despite the fact that (or rather precisely
because) it is consistent with zero.  At the $2\,\sigma$ level, this 
constrains the scaleheight to be $\kappa^{-1}>7\,$kpc.  If the local
halo were composed of two components, one highly flattened and one roughly
round, then one would expect the density to fall off locally over distances
that are short compared 7 kpc.  Hence, this result should help constrain
2-component halo models.

\subsection{Color-Magnitude Relation
\label{sec:cmr}}

I fit for a color-magnitude relation of the form
$M_V = a(V-J)_0 + b$ (eq.\ [\ref{eqn:cmr}]) and find
\begin{equation}
a = 3.59,\qquad b=0.69.
\label{eqn:abeval}
\end{equation}
The formal uncertainties on these parameters are very small, of order
0.01 mag.  However, recall from the discussion above equation 
(\ref{eqn:gpu2}) that $b$ is completely degenerate with the amplitude
of the velocity ellipsoid, which was fixed for purposes of the fit but which
actually has a statistical uncertainty of 6\% (and a comparable systematic
error).  Hence, the true error in $b$ is about 0.2 mag.  The total
error in $a$ is probably not much larger than the formal error.

\subsection{Luminosity Function
\label{sec:lf}}

	The LF is parameterized by 13 separate 1-mag bins, with
centers from $V=3$ to $V=15$.  I find (as did \citealt{gbf} when they
studied the disk LF) that ML estimates of the LF
tend to magnify Poisson fluctuations according to the following mechanism.
First suppose that the true LF has a dip at a certain bin.  Observational
errors will scatter stars from the two neighboring bins into this bin,
thus tending to wash out the dip. Hence, a ML fit, when
confronted by a dip in the observed distribution will tend to accentuate 
it as it reconstructs the underlying (true) LF. Now suppose
that the true LF is flat over three adjacent bins but because
of Poisson fluctuations the central bin is depressed.  ML
will also accentuate this dip in an attempt to reconstruct the ``true''
LF.  Hence, particularly for bins with low total counts, ML
can introduce structure that is not really present.  I handle this
potential problem by 
imposing a ``roughness'' penalty $\Delta {\cal L} = 16$(difference/sum)$^2$,
where ``sum'' and ``difference'' refer to the sum and difference of the
LF in each pair of neighboring LF bins.  This is equivalent to imposing
a $\Delta\chi^2=1$ penalty when neighboring bins differ by 35\%.
Thus, if the data really demand a steep gradient, this penalty will permit
one, but it will squash spurious gradients.

Figure \ref{fig:one} shows the resulting LF.  This LF is significantly
correlated with the completeness parameters (see eq.\ [\ref{eqn:complete}]),
which are derived simultaneously,
\begin{equation}
f_\break = 43\pm 6\%,\quad V_\break=18.27\pm 0.04.
\label{eqn:compfac}
\end{equation}
In the Appendix, I consider arguments that might lead one to suspect that this 
estimate of $f_\break$ could be substantially too low.  I find that these 
arguments are not compelling and therefore adopt the LF calculated using
equation (\ref{eqn:compfac}) as the best estimate.  Nevertheless, in order
to gain a sense of the possible role of such a systematic effect, I
also show in Figure \ref{fig:one} the LF under the assumption that
$f_\break=65\%$, the highest value that I consider to be plausible.

Figure \ref{fig:two}
compares the derived LF (with $f_\break=43\%$) 
to those of several
previous measurements of the halo LF: those of \citet{bc}, \citet{dahn},
and \citet{gfb}, which were all previously compared by \citet{gbf}.
As explained there, the first two LFs are labelled ``BC/CRB'' and
``DLHG/CRB'' to indicate that they have been corrected from the 
originally published LFs to take account of the kinematic selection
using the velocity ellipsoid of \citet{crb}, which is very similar
to the ellipsoid of \citet{gp}, and to the one derived for the present
sample.  The present measurement is in reasonably good agreement with
the DLHB/CRB determination over the range $9\leq M_V\leq 14$
covered by the latter.  It disagrees strongly with both the BC/CRB and
\citet{gfb} determinations (which are in good agreement with each other).
The new measurement extends over a much wider magnitude range and
has substantially smaller error bars than any previous determination.

	The LF evaluations at $M_V=15$, and to a lesser extent at
$M_V=14$, should be interpreted cautiously because they depend sensitively
on model assumptions.  To understand this point, one should consider how
ML ``thinks'' when fitting the LF.  To zeroth order, it forms an LF in
the naive way: by counting the number of stars whose dereddened observed 
$(V-J)_0$ color and the color-magnitude relation put them in a corresponding
$M_V$ bin, and dividing this number
by the total effective volume probed by the survey
for stars of that $M_V$.
There are respectively 40 and 18 stars in the final two bins, but
only roughly 20 and 6 of these are assigned to these bins by the final ML LF 
fit shown in Figures \ref{fig:one} and \ref{fig:two}.  
What prevents ML from assigning
much higher densities to the LF at these faint magnitudes?  In the next
brightest bin there are 159 stars.  This is about four times larger than in
the $M_V=14$ bin despite the fact that the LF is roughly the same because
the effective volume grows rapidly with luminosity at these faint
magnitudes.  As described above, ML takes this as a zeroth-order estimate
for the number expected in this bin.  It then considers how many of these
are expected to scatter into neighboring bins because the color errors
$\sigma(V-J)\simeq\sigma(V)=0.25$ induce errors in $M_V$ of 
$a\sigma(V)\sim 0.9$.  That is, roughly 15\% of these 159 stars are expected
to scatter into the $M_V=14$ bin and a few percent into the $M_V=15$ bin.
It is by accounting for this scatter, as well as scatter form brighter
bins, that ML achieves its final estimate.  This estimate therefore depends
quite senstively on the adopted value of $\sigma(V)$, which is described
very simply in the model, but could in principle actually be a function of $V$
or of other variables.  

	In addition, because of the very small number of
detected stars in these final bins, there is a potential problem of 
contamination from non-halo stars.  Contamination is not generally a problem
because, as I argued in \S~\ref{sec:selcrit}, the discriminator $\eta$
is limited to regions well away from main-sequence stars and white dwarfs.
However, the density of the very dim stars on the RPM diagram is extremely
low (see Fig.\ 3 from \citealt{faint}), so even the low residual level
of contamination could play a role.

	To establish the LF at $M_V\geq 14$ more securely, and
since the total number of stars in the last two bins is very small
(58), the simplest approach would be to obtain $V$-band photometry
for all of them.  If the ML result is correct, the majority of these will
be found to have scattered in from brighter bins.  Metallicities and radial
velocities from spectra of the truly red stars could then resolve issues of
contamination.  The LF of the final two bins would then rest on much firmer
ground.

\section{Discussion
\label{sec:discuss}}

To high precision (roughly $2\,\kms$) the LSR is not moving with respect
to the halo in either the radial or vertical directions.  If the halo
itself has no radial motion, the first result sharply contradicts
the conclusion of \citet{bs} based on gas motions that the LSR is moving 
outward at $14\,\kms$.  On the other hand, it is reasonably consistent with
the radial-motion estimate of \citet{ms} based on carbon stars.  More
specifically, I find that the halo is moving at $11.4\pm 2.2\,\kms$
relative to the Sun, and they find that the outer-Galaxy carbon stars
are moving at $15.6\pm 1.7\,\kms$.

I find that all three off-diagonal components of the velocity
dispersion tensor are small, within $\sim 2\%$ of zero.  
The only previous measurements of these quantities (PG$^3$) were
consistent with zero, but with errors that were about 4 times 
larger.  To date, I am not aware of any effort to
predict the off-diagonal terms from theory.

The measurement presented here of the LF confirms the basic peaked shape 
found by 
\citet{dahn}, but with about 40 times more stars and therefore covering a
magnitude interval that is roughly twice as large.  It is inconsistent with
the flat LF found by \citet{bc}, and \citet{gfb} (although in principle,
since the latter determination was based on stars away from the solar
neighborhood, it cannot be rigorously ruled out by my measurement).
The present measurement is in rough agreement with that of \cite{bc} at
brighter magnitudes, $M_V<9$.

A shortcoming of the present approach is that there is no information about 
distances within the data set, so the scale of the velocity ellipsoid
must be set by external information.  The distance scale could be set
by obtaining either radial velocities (RVs) or trig parallaxes for a 
{\it representative} (i.e., random) subset of the stars in the sample.  
The former would yield a statistical parallax solution.
I stress ``representative'' because if the subsample is biased, for
example is weighted toward stars with extreme kinematics and/or
low metallicities, then the scale of the velocity ellipsoid will be
overestimated by statistical parallax because the stars with RVs 
move faster than the those in the sample as a whole.
It would be misestimated by trig parallax both because the selected
stars would be faster and subluminous compared to the sample as a whole.
Hence, one must choose a fair sample, and then make use of archival data
only for stars within that sample.

 From the standpoint of maximizing the precision of the distance-scale
measurement with the minimum effort, statistical parallax is to be much
preferred over trig parallax.  Even velocity errors of 
$\sim 20\,\kms$ are quite adequate for a statistical parallax measurement
with the limiting precision $\sigma(\eta)/\eta = 0.65 N^{-1/2}$
\citep{pg1}.  Here $N$ is the number of stars in the statistical parallax
sample and $\eta$ is the distance-scale parameter.  Even assuming 
perfect parallaxes, the limit for the trig parallax technique is
$\sigma(\eta)/\eta=0.2\ln 10\sigma(M_V)N^{-1/2}$.  
Given the large number of halo stars in
the Revised NLTT Catalog and the relative ease  of making RV compared to trig
parallax measurements, the modest per-star advantage of trig parallax
will be overwhelmed by the mass-production techniques available
for RVs.  However, good trig parallaxes would provide information on the
luminosity of individual stars, which cannot be obtained from statistical
parallax techniques.  Thus, the two approaches are complementary.

\acknowledgments 
I thank Samir Salim for invaluable discussions about both the content
and presentation of this paper.
Work supported by JPL contract 1226901.

\appendix
\section{Completeness
\label{sec:completeness}}

\begin{figure}
\plotone{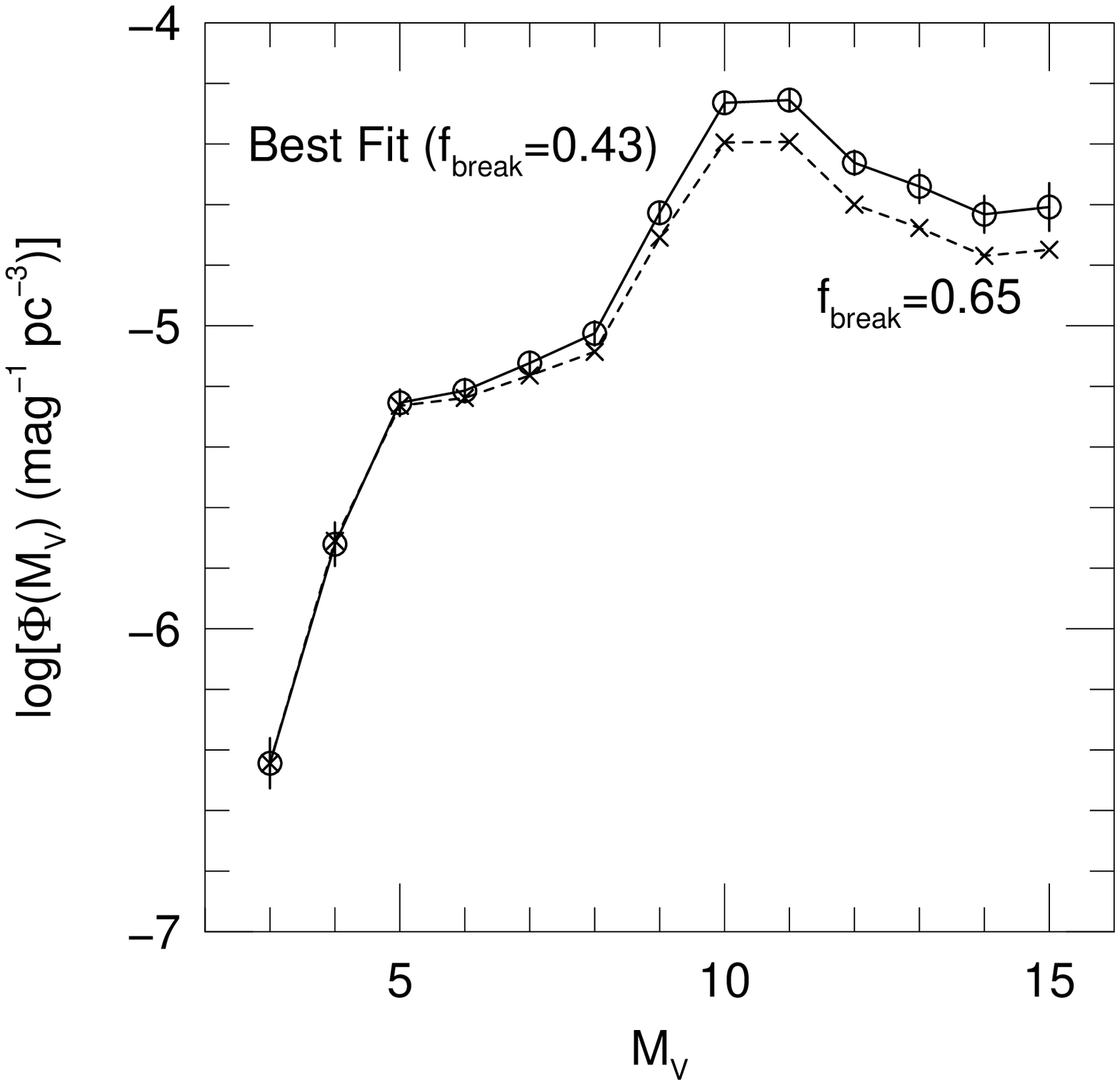}
\caption{\label{fig:one}
Logarithm of the luminosity function (LF) derived from 4588 subdwarfs from the
Revised NLTT Catalog \citep{bright,faint}.  The solid curve and
open symbols with error bars represent the best fit.  The
dashed curve and crosses represent the fit under the assumption
that the derived catalog completeness has been seriously underestimated.
The main features of the LF remain the same.  The faintest two
bins should be interpreted cautiously (see \S~\ref{sec:lf}).
}\end{figure}

\begin{figure}
\plotone{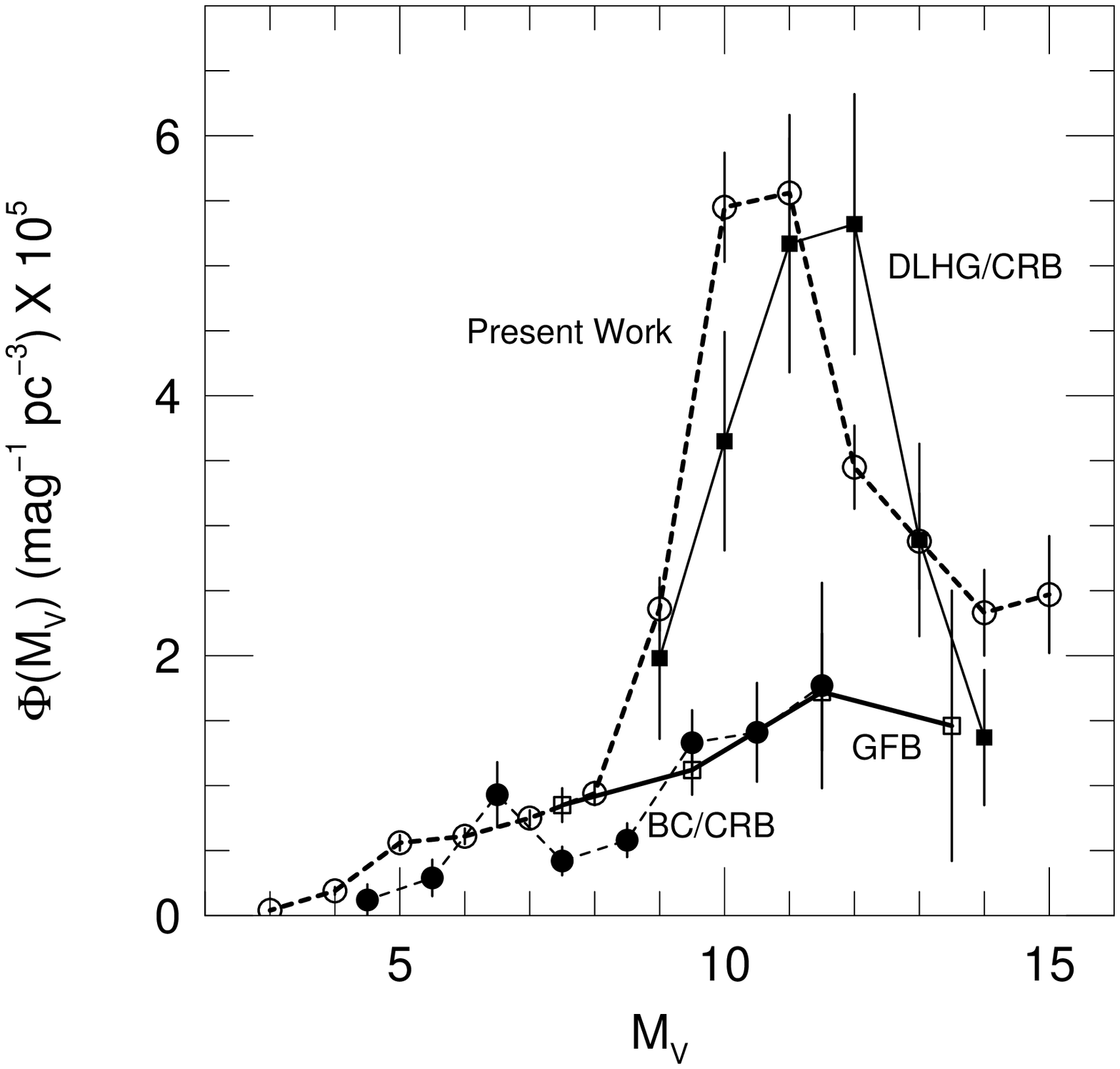}
\caption{\label{fig:two}
Comparison of four halo LFs: The original determinations by
\citet{dahn} (DLHG) and \citet{bc} (BC) have been rescaled by 
\citet{gfb} using the velocity ellipsoid of \citet{crb} (CRB).
The present work confirms the ``bump'' in the LF found by
DLHG at $M_V\sim 11$ as well as the fall-off toward brighter
mags found by \citet{bc}, but with much smaller error bars
in both cases.
}\end{figure}

The principal source of systematic errors in this analysis is incompleteness
(or rather, possible misestimation of the completeness) of the Revised
NLTT Catalog.  As shown in Figure \ref{fig:one}, 
such misestimation can significantly affect the determination of the LF. 
However, I find that it does not affect the estimate of the velocity 
ellipsoid.

	In principle, catalog completeness could be a function
of all six observables, i.e., the $V$ and $J$ magnitudes, the proper motion,
and the position on the sky.  While the {\it relative completeness} of the
Revised NLTT compared to the original NLTT is very well understood
\citep{faint}, there is substantially less information available about the
{\it absolute completeness} of the underlying NLTT.  

The completeness of NLTT can be tested either externally or internally.
An external check requires an independent search for high proper-motion
stars, either over the whole sky or some fraction of it.  By comparing
to the {\it Hipparcos} \citep{hip} and Tycho-2 \citep{t2} catalogs, 
\citet{bright}
concluded that NLTT is nearly 100\% complete for $V\la 11$ and for
Galactic latitudes $|b|>15^\circ$.  Even near to the plane, completeness
is close to 100\% for $\mu>400\,\masyr$.  \citet{monet} searched for high
proper-motion ($\mu>400\,\masyr$) 
stars to faint magnitudes toward 1378 deg$^2$,
and found 241 stars, only 17 of which (their Tables 2 and 3) they could
not match to NLTT.  In fact, two of these 17 are actually NLTT 
stars, namely 58785 and 52890,
which correspond to entries 1 and 8 from Table 3.  Thus, over these surveyed
areas, NLTT is $94\%\pm 2\%$ complete for $\mu>400\,\masyr$.  However,
the {\it Luyten Half-Second} (LHS) catalog \citep{lhs}, a subset of NLTT
which actually extends somewhat below $500\,\masyr$, is almost certaintly 
substantially more complete than is the NLTT at lower proper motions, so this 
measurement cannot be regarded as representative of NLTT as a whole.  

Unfortunately,
there are no sytematic studies comparing NLTT detections with an independent
search for proper motion stars in the range $180\,\masyr<\mu<400\,\masyr$
and at faint magnitudes.  In the absence of such {\it external} checks, 
\citet{flynn} conducted an {\it internal} completeness determination,
whose approach is very closely related to the completeness measurement
carried out here by ML.  For a complete sample drawn from
a stellar population that is uniformly distributed in space, the number
of stars in the phase-space volume $[\mu_1,\mu_2]\times [V_1,V_2]$, should
be (up to Poisson statistics) exactly 8 times the number in the volume
$[2\mu_1,2\mu_2]\times [V_1-5\log 2,V_2-5\log 2]$.  This is because
the former physical volume is 8 times larger while the physical velocities
being probed are exactly the same.  \citet{flynn} made a series of such
comparisons at 0.5-magnitude intervals, and by multiplying these together,
found that the completeness at $R_\nltt=18.5$ is 65\% that at
$R_\nltt=13$ (which latter they assumed to be 100\%).

	\citet{monet} pointed out that {\it any} effect that reduced
detections of more distant, slower moving stars (relative to faster nearby
ones) could masquerade as incompleteness of fainter (relative to brighter)
stars under this test.  In particular, they argued that more distant stars
would, being on average farther from the plane, have reduced density.
Even a small difference of 4\% in mean density per comparison could
lead to the ``observed'' $f_\break=65\%$ completeness at 
$R_{\nltt,\break}=18.5$, since it would be multiplied together
11 times between $R_\nltt=13$ and 18.5.  That is,
$(1- 0.04)^{11}\sim 65\%$. \citet{monet} presented a figure showing
that when the procedure is carried out on stars at lower Galactic latitude,
the effect is much reduced.

	This critique is important because the 2-parameter characterization
of incompleteness that I use here (see \S~\ref{sec:modparms}) in essence
embodies the \citet{flynn} method.  As noted in \S~\ref{sec:lf},
I find $f_\break = 43\% \pm 6\%$ at $V_\break=18.27$.
This fraction is consistent with the \citet{flynn} result at the 
$1.5\,\sigma$ level when account is taken of the fact that I have assumed
100\% completeness at $V=12$ rather than $R_\nltt=13$.  The magnitude 
$V_\break$ is slightly brighter than the value found by \citet{flynn}, but
this is to be expected, since the Revised NLTT Catalog depends on 2MASS 
$J$-band cross identification.  

	However, the \citet{monet} critique cannot account
for the result found in equation (\ref{eqn:compfac}) for several
reasons.  First, one does not expect the density of halo subdwarfs to fall 
off significantly within the volume probed by NLTT.  Second, the ML fit
actually allows for such a fall-off, so that even if the actual density
did not satisfy this theoretical expectation, the fit would automatically
take account of the fall-off.  (In fact according to eq.\
[\ref{eqn:nukappa}] the best fit
is consistent with uniform density.)\ \  Third, \citet{flynn} 
are unable to reproduce
the \citet{monet} figure 3 and instead find relatively comparable results
when they apply their test to high-latitude and low-latitude stars.

	Nonetheless, it remains possible that there is some {\it other}
effect that reduces the counts of distant relative to nearby stars of
the same class.  One such effect would be incompleteness as a function
of proper motion, rather than magnitude.  Such incompleteness could be
very pernicious because if the more distant (and hence slower moving)
stars were under-represented by a mere 4\%, then (as outlined above)
the effect would be exponentiated and would generate a large apparent 
incompleteness as a function of magnitude.  I therefore tested this
hypothesis by including a proper-motion term for the completeness within
the ML fit.  However, this term did not improve the fit even slightly.
One is thus left without any plausible explanation for the relative lack
of more distant, slower stars, other than incompleteness as a function
of magnitude.

The little direct information we have on the completeness of NLTT
at faint magnitudes and relatively low proper motions is reasonably
consistent with the ML estimates of completeness derived here.
\citet{nreid} searched for high proper motion stars in a single Schmidt 
field toward
the north Galactic pole.  He recovered 63 stars from NLTT as well as
15 stars that met NLTT selection criteria but were not in NLTT.  This
appears to correspond to a mean completeness of 81\%. 
However, three of the 15 have proper motions
$180\,\masyr<\mu<200\,\masyr$, i.e., within $1\,\sigma$ \citep{faint}
of the NLTT limit.  Stars that scatter across the selection boundary
are {\it already} taken into account in the modeling procedure and should
not be counted as due to ``incompleteness''.  Hence, the true mean completeness
of NLTT in the \citet{nreid} survey area is more like $84\pm 5\%$.   If one 
restricts consideration to stars $V<V_\break$ (beyond which the Revised NLTT  
Catalog is more sensitive to the incompleteness of 2MASS than NLTT),
then $f_\break=43\pm 6\%$ corresponds to a mean completeness of halo stars
of $78\pm 3\%$, in $1\,\sigma$ agreement with the value just derived
from the \citet{nreid} study.  

In brief, there is no strong evidence to challenge the completeness
estimate given by the ML fit.


\clearpage

\end{document}